\documentclass[prl,twocolumn,superscriptaddress,preprintnumbers,amsmath,amssymb]{revtex4-1}
\usepackage{graphicx}
\usepackage{color}
\usepackage{bm}
\usepackage{ulem} 

\begin{document}
\title{
Maxwell's Demon Based on a Single Qubit}
\author{J.P. Pekola}
\affiliation{Low Temperature Laboratory, Department of Applied Physics, Aalto University School of Science, P.O. Box 13500, 00076 Aalto, Finland}
\author{D.S. Golubev}
\affiliation{Low Temperature Laboratory, Department of Applied Physics, Aalto University School of Science, P.O. Box 13500, 00076 Aalto, Finland}
\author{D.V. Averin}
\affiliation{Department of Physics and Astronomy, Stony Brook University, SUNY, Stony Brook, NY 11794-380, USA}

\begin{abstract}
We propose and analyze Maxwell's demon based on a single qubit with avoided level crossing. Its operation cycle consists of adiabatic drive to the point of minimum energy separation, measurement of the qubit state, and conditional feedback. We show that the heat extracted from the bath at temperature $T$ can ideally approach the Landauer limit of $k_BT\ln 2$ per cycle even in the quantum regime. Practical demon efficiency is limited by the interplay of Landau-Zener transitions and coupling to the bath. We suggest that an experimental demonstration of the demon is fully feasible using one of the standard superconducting qubits.
\end{abstract}

\date{\today}

\maketitle
\begin{figure}
  \begin{center}
    \includegraphics[scale=.18]{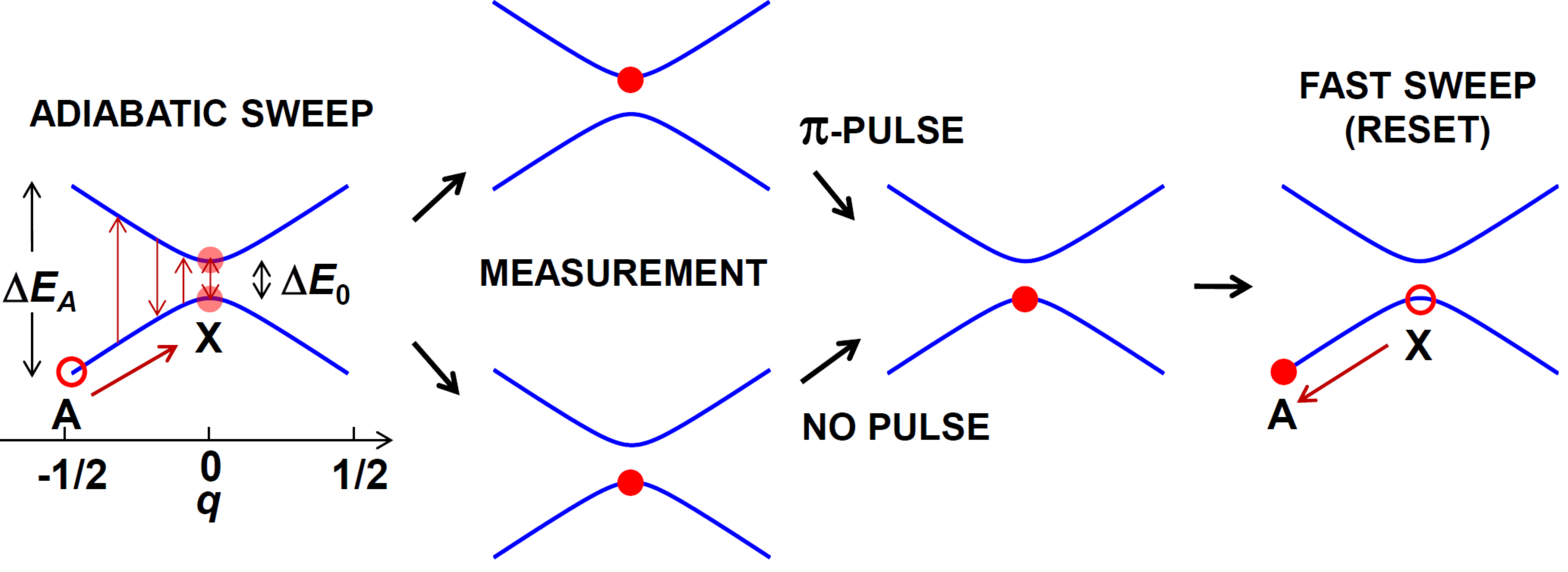}
    \end{center}
    \caption{The scheme and operation cycle of the qubit demon.}
    \label{fig1}
\end{figure}

Controllable small systems, e.g. in form of molecules, micro-beads or nanoelectronic circuits \cite{bustamante05,batalhao14,ciliberto10,blickle06,kung12,saira12}, have recently made it possible to experiment on thermodynamics in the regime where fluctuations play an important role. One of the most interesting issues in such stochastic thermodynamics is the role of information. On the level of thought experiments this discussion dates back all the way to Maxwell's demon (MD) \cite{leff} and then to Landauer's principle \cite{landauer61}. Till now, majority of the studies, both theoretically and in particular experimentally \cite{toyabe10,berut12,koski14a,roldan14}, have been focusing on the classical regime; see, however, \cite{lloyd97,quan06,horowitz14,brandner15,parrondo15} and references therein for theories on quantum systems. The aim of this Letter is to present a simple and experimentally feasible quantum MD that can operate at the limit of thermodynamic efficiency. Demon operation is based on a qubit, a two-level system (TLS) with avoided level crossing between the ground ($g$) and the excited ($e$) state as illustrated in Fig. \ref{fig1}. The TLS is coupled weakly to a bath at temperature $T$. The MD moves around the level crossing by tuning the control parameter $q$.

We first present an ideal operation cycle and results for the MD without justifying their feasibility. The cycle start at point A in Fig. \ref{fig1}, where the level separation $\Delta E$ is maximum denoted by $\Delta E_A$. If $\Delta E_A\gg k_BT$, the system is in the ground state with zero entropy. In the first leg, $q$ is moved adiabatically to the point X with minimum level spacing $\Delta E_0$. We assume that the ramps are linear in time, i.e. $\dot q = {\rm constant}$ within each leg. Ideally, for $\Delta E_0 \ll k_BT$ and fully adiabatical ramp, the entropy of the TLS increases by $k_B \ln 2$, corresponding to heat extracted from the bath equal to Landauer value $Q_L=k_BT\ln 2$. This part of the cycle is equivalent to that in a classical TLS -- see Ref.~\cite{koski14a}, although the non-vanishing $\Delta E_0$ creates essential differences in the quantum regime. On reaching point X, a measurement of the qubit state is performed. If the detector measures the appropriate quantity: the energy or the curvature $d^2E/dq^2$ (which, depending on the physical realization of the system can represent, e.g., effective capacitance or inductance in an electric circuit) it does not perturb this state and therefore, does not incur any thermodynamic costs. Note that, in contrast to a classical MD, the possibility to realize such a measurement required for the thermodynamically efficient quantum MD, is not automatic, but relies on the feature of adiabatic dynamics to leave the qubit density matrix diagonal in the energy basis. After the measurement, the detector, ideally, ``knows'' with certainty whether the system is in the ground state or in the excited state in each cycle. In the repeated cycles, the two states are realized with equilibrium thermal probabilities. This makes it possible to apply conditional feedback: if the system is in the ground state, the control gate is moved quickly back to the original position. In the excited state, a $\pi$-pulse is applied to the qubit swapping the state $e\rightarrow g$, the process in which the work done by the source is $-\Delta E_0$ \cite{talkner07,esposito09}, followed again by a fast ramp in the ground state back to the original position A. Extracted heat, averaged over several such ideal cycles is
\begin{equation} \label{qd4}
\langle Q \rangle  = -\beta^{-1}\ln (1+e^{\beta \Delta E_0})+\frac{\Delta E_0}{1+e^{-\beta \Delta E_0}} \, ,
\end{equation}
where $\beta =(k_BT)^{-1}$.

Several processes can compromise the ideal performance of the qubit demon. First, one may extract less heat from the bath if the ramp A $\rightarrow$ X is not fully adiabatic. Also, the feedback ramp X $\rightarrow$ A needs to be ``fast'' to avoid bath-induced transitions but ``slow'' to avoid Landau-Zener excitation out of the ground state. Both processes would create extra dissipation. To consider them quantitatively, we start with the operator of the power injected into the system, $P(t)=\dot H(t)$, where $H(t)$ is the system Hamiltonian. Work $W$ done on the system over a time interval $\tau$ is then $W=\int_0^\tau P(t) dt$. Since we are mainly interested in quantities averaged over many cycles, it is convenient to describe the dynamics with the density matrix $\rho(t)$. For average work
$\langle W\rangle = \langle \int_0^\tau P(t) dt \rangle$ one finds by elementary analysis \cite{solinas13}
\begin{equation} \label{na1}
\langle W\rangle =\langle H(\tau)\rangle -\langle H(0)\rangle -\int_0^\tau  {\rm Tr} \big [\dot \rho (t) H(t)\big ]  dt.
\end{equation}
Equation \eqref{na1} represents the first law of thermodynamics, combining the change of the internal energy and the average
heat generated in the bath:
$\langle Q\rangle =   -\int_0^\tau  {\rm Tr} \big [\dot \rho (t) H(t)\big ]  dt$.
In the basis $\{|g\rangle,|e\rangle\}$ of instantaneous eigenstates of the qubit, this expression gives
\begin{equation} \label{na4}
\langle Q\rangle =  \int_0^\tau dt \dot\rho_{gg}(t)\Delta E(t).
\end{equation}

The standard master equation for the qubit density matrix $\rho(t)$, ignoring pure dephasing, is
\begin{eqnarray} \label{na5}
&&\dot \rho_{gg}=-\Omega \Re{\rm e}[\rho_{ge}e^{i\int_0^t \Delta E(t')dt'/\hbar}]-\Gamma_\Sigma \rho_{gg} +\Gamma_\downarrow \nonumber \\
&& \dot \rho_{ge} = \Omega(\rho_{gg}-1/2)e^{-i\int_0^t \Delta E(t')dt'/\hbar} -\frac{1}{2}\Gamma_\Sigma \rho_{ge}\, ,
\end{eqnarray}
where $\Omega =\frac{\epsilon}{q^2+\epsilon^2}\dot q$ depends on the ramp rate of $q$, $\epsilon = \Delta E_0/(2\Delta E_A)$, and the level spacing is $\Delta E (t) = 2\Delta E_A\sqrt{q(t)^2+\epsilon^2}$. Point A corresponds to $q=-1/2$, whereas point X is at $q=0$. The relaxation
and excitation rates of the qubit are, respectively,
\begin{equation} \label{qd7a}
\Gamma_\downarrow (t) = \frac{g^2}{\hbar} \frac{\Delta E(t)}{1-e^{-\beta \Delta E (t)}},\,\,\,\,\Gamma_\uparrow (t) = \frac{g^2}{\hbar} \frac{\Delta E(t)}{e^{\beta \Delta E (t)}-1}\, ,
\end{equation}
and $\Gamma_\Sigma = \Gamma_\downarrow + \Gamma_\uparrow$. Parameter $g^2$ denotes the dimensionless strength of coupling to the bath.

To learn about stochastics beyond the averages above, we analyze quantum trajectories corresponding to the master equation \eqref{na5} for the evolution of the wave function $|\psi (t) \rangle = a(t)|g \rangle + b(t)|e\rangle$. According to the quantum jump approach \cite{dalibard92}, the probability to relax to $|g\rangle$ (get excited to $|e\rangle$) over a short time interval $\Delta t$ is given by $\Delta p_\downarrow = \Gamma_\downarrow |b(t)|^2 \Delta t$ ($\Delta p_\uparrow = \Gamma_\uparrow |a(t)|^2 \Delta t$). If such a jump does not occur, the system evolves as
\begin{eqnarray} \label{qd7}
&&\dot a= -\frac{1}{2}\Omega e^{-i\int_0^t\Delta E(t')dt'/\hbar} b +\frac{1}{2}\Delta \Gamma a|b|^2 \nonumber \\
&&\dot b= +\frac{1}{2}\Omega e^{+i\int_0^t\Delta E(t')dt'/\hbar} a -\frac{1}{2}\Delta \Gamma b|a|^2,
\end{eqnarray}
where $\Delta \Gamma (t) = \Gamma_\downarrow (t) - \Gamma_\uparrow (t) = g^2 \Delta E(t)/\hbar$. The evolution of the system is obtained numerically using a stochastic simulation based on these equations. The heat dissipated into the bath is given by $(-)\Delta E$ when the system relaxes (gets excited) \cite{hekking13}, see also \cite{horowitz13}.

We analyze the MD cycle and errors quantitatively assuming that the measurement and the conditional $\pi$-pulse are fast enough (duration $\ll \Gamma_\Sigma^{-1}$) so that no heat is exchanged in this time interval. Consider first the quasi-static sweep A $\rightarrow$ X in
time $\tau=(2\dot q)^{-1}$. For $\tau \rightarrow \infty$, the population obeys
\begin{equation} \label{qd2}
\rho_{gg}(t)=\Gamma_\downarrow/\Gamma_\Sigma = (1+e^{-\beta\Delta E(t)})^{-1}
\end{equation}
according to Eq.~\eqref{na5}. Inserting this expression into Eq.~\eqref{na4}, we obtain
\begin{equation} \label{qd3}
\langle Q\rangle =  -\beta^{-1}\ln\Big(\frac{1+e^{\beta\Delta E_0}}{1+e^{\beta\Delta E_A}}\Big)+\frac{\Delta E_0}{1+e^{-\beta\Delta E_0}}-\frac{\Delta E_A}{1+e^{-\beta\Delta E_A}}.
\end{equation}
Assuming everywhere below that $\beta \Delta E_A \gg 1$, we recover Eq.~\eqref{qd4}. If $\Delta E_0=0$, we obtain then the most
familiar result $\langle Q\rangle =-k_BT\ln 2=-Q_L$. For such a degenerate case, the average heat equals the heat obtained in each
adiabatic realization, but this is not so for $\Delta E_0> 0$.

\begin{figure}
    \begin{center}
    \includegraphics[scale=.2]{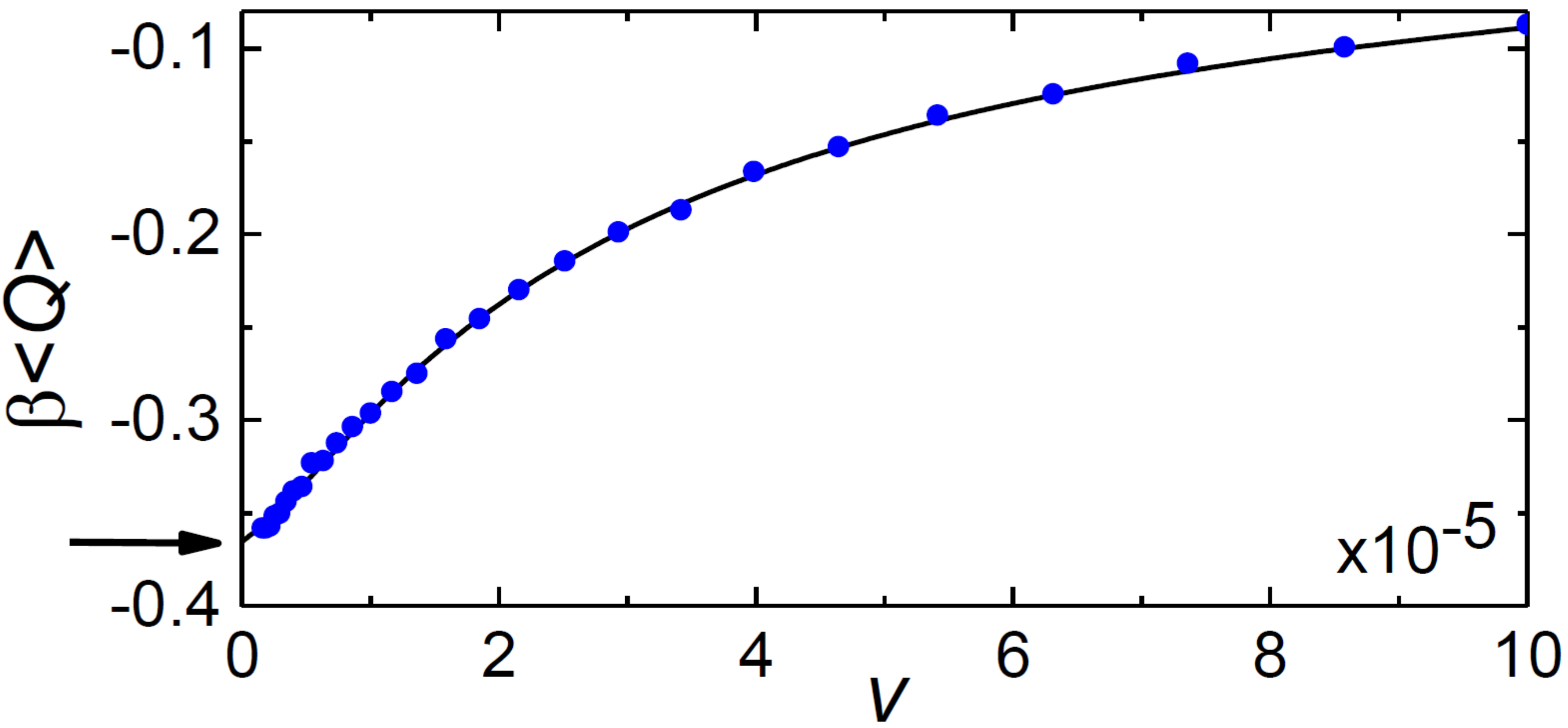}
    \end{center}
    \caption{$\langle Q\rangle$ as a function of the rate $v=\hbar \dot q/\Delta E_A$ in a quasi-static A $\rightarrow$ X ramp. The horizontal arrow points the result of fully adiabatic evolution based on Eq.~\eqref{qd4}. The solid black line is the full numerical solution of the master equation \eqref{na5}. The blue dots are based on the stochastic quantum trajectory simulations. The parameters are $\epsilon = 0.1$, $\beta\Delta E_A = 10$, and $g^2 = 10^{-3}$.}
    \label{nonad}
\end{figure}

For finite values of $\tau$, the evolution becomes non-adiabatic and can be analyzed using Eq.~\eqref{na5}. Provided
$\hbar/(2\pi \epsilon^2\Delta E_A\tau)\ll 1$, one can neglect Zener tunneling processes and set $\Omega=0$ in Eq.~\eqref{na5}.
Then one finds exactly
\begin{equation} \label{na8}
\langle Q\rangle = \int_0^\tau dt\int_0^{t} dt'
\frac{\beta\Gamma_\Sigma(t)\Delta E(t)\Delta\dot E(t')e^{-\int_{t'}^{t} ds\Gamma_\Sigma(s)}}{4\cosh^2[\beta\Delta E(t')/2]}.
\end{equation}
In Fig.~\ref{nonad}, we present results on $\langle Q \rangle$ normalized by $k_BT$ as a function of the ramping rate $v= \hbar/(2\tau\Delta E_A)$ in the A $\rightarrow$ X ramp for chosen parameters. The black line is based on the master equation \eqref{na5} together with \eqref{na4}, and is indistinguishable from Eq.~\eqref{na8}. The blue symbols are from the stochastic simulation with $50000$ repetitions for each point. The horizontal arrow indicates the adiabatic result of Eq.~\eqref{qd4}.

The errors in the fast return ramp X $\rightarrow$ A in time $\tau_R$ are produced by the Landau-Zener processes which generate finite population of the excited state. If the excited state amplitude is $b_f$ at the end of the sweep, there is ``quantum'' heat $\sim |b_f|^2 \Delta E_A$ dissipated into the bath in the next adiabatic ramp to $q=0$. On the other hand, if the sweep is too slow and $\Gamma_\uparrow \tau_R \ll 1$ does not hold, the system can be excited by the bath, and there is extra ``classical'' generated heat when the system relaxes either before or during the adiabatic ramp in the next cycle. These two errors constrict the ramp time to the interval $\hbar/(\epsilon^2 \Delta E_A)\ll \tau_R \ll \Gamma_\uparrow ^{-1}$ as a condition of approaching the ideal operation. The lower bound is the same as for the standard Landau-Zener dynamics.

We estimate these two contributions to heat as follows. For weak dissipation and system being mostly in the ground state, we write the evolution of $b$ as (see \eqref{qd7})
$\dot b \approx \frac{1}{2}\Omega e^{+i\int_0^t \Delta E(t')dt'/\hbar}$.
Linearizing the expression in the exponent with respect to $q$, 
and denoting the dimensionless rate as $v_R =\hbar /(2\tau_R \Delta E_A)$, we obtain the final population at the
end of the ramp as
\begin{equation} \label{qd10}
|b_f|^2 \approx v_R^2/(16\epsilon^4)\, ,
\end{equation}
valid for $|b_f|^2 \ll 1$. This result agrees with the one obtained in \cite{zurek06}.

The corresponding classical error is determined by the probability $p_\uparrow$ of a thermal excitation $g \rightarrow e$ during the ramp, again assuming that $|a|^2\approx 1$. This yields
\begin{equation} \label{qd11}
p_\uparrow \approx \int_0^\infty dt' \Gamma_\uparrow (t')\approx  \frac{\sqrt{\pi} g^2 \epsilon^{3/2}e^{-2\epsilon \beta \Delta E_A}}{\sqrt{\beta\Delta E_A}}v_R^{-1}.
\end{equation}
Both excitation processes, described by Eqs.~\eqref{qd10} and \eqref{qd11}, combine to produce extra average heat dissipated in a cycle:
$\langle Q_{\rm diss}\rangle \approx (|b_f|^2 + p_\uparrow)\Delta E_A$.
Minimum of $\langle Q_{\rm diss}\rangle$,
\begin{equation} \label{minQ}
\langle Q_{\rm diss}\rangle_{\rm min}=\beta^{-1}[\frac{27\pi}{64}\frac{(g^2 \beta \Delta E_A e^{-2\epsilon \beta \Delta E_A})^2}{\epsilon}]^{1/3}
\end{equation}
with respect to $v_R$ is then obtained at
\begin{equation} \label{qd15}
v_{R,\rm opt} = (\frac{8\sqrt{\pi} g^2 \epsilon^{11/2}e^{-2\epsilon \beta \Delta E_A}}{\sqrt{\beta \Delta E_A}})^{1/3}.
\end{equation}

In this optimization, we considered only the most straightforward linear return ramp, for which the excitation amplitude $b_f$ \eqref{qd10} decreases as $1/\tau_R$. In principle, $b_f$ can be suppressed more strongly if the return ramp is switched not abruptly, as in the cycles considered above, but smoothly. As discussed in the Supplementary Material \cite{suppl}, the amplitude $b_f$ decreases as $(\hbar /\Delta E_0 \tau_R)^n$, if 
only the $n$th-order derivative of the switching function is discontinuous, while all previous derivatives are continuous, and exponentially, 
if the switching function has continuous derivatives of any order.

\begin{figure}
    \begin{center}
    \includegraphics[scale=.18]{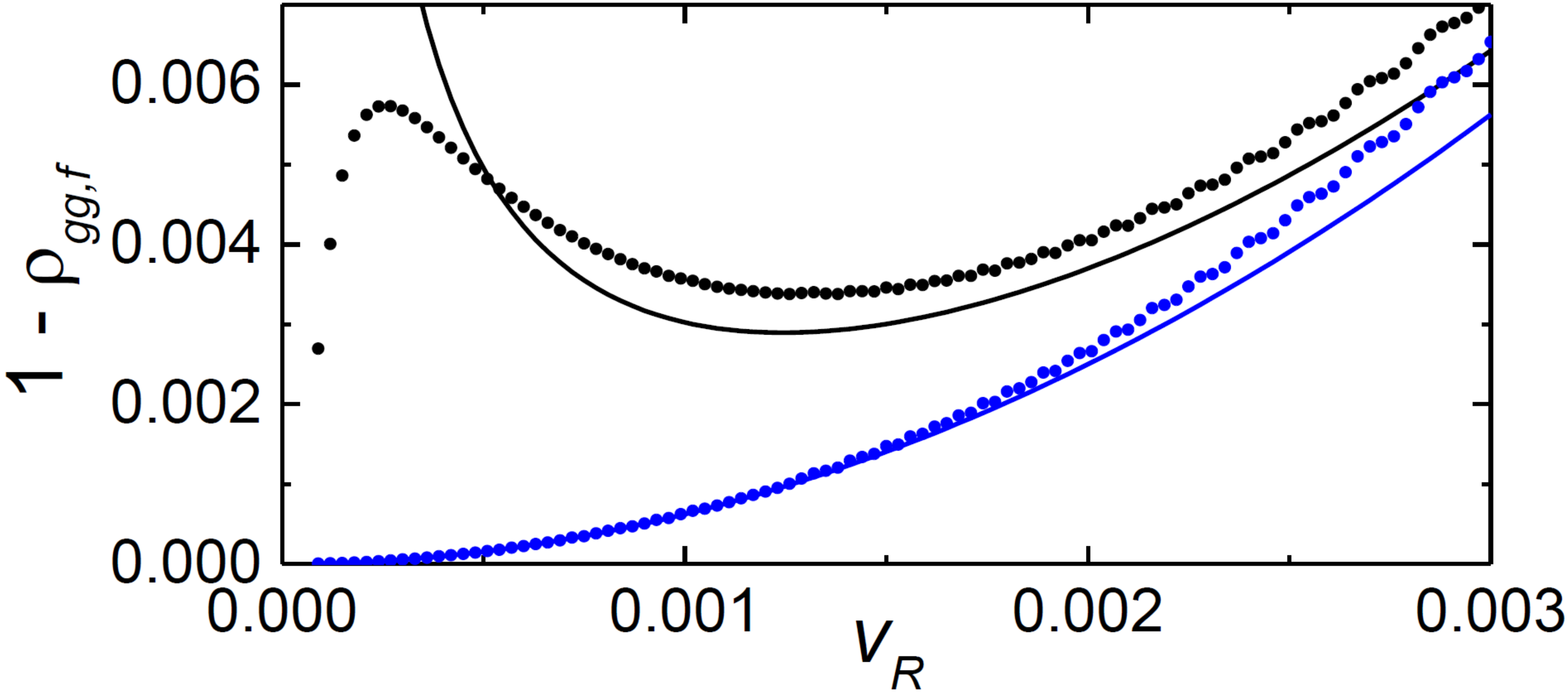}
    \end{center}
    \caption{Population of the excited state at the end of the return sweep X $\rightarrow$ A based on the master equation \eqref{na5} (symbols), together with the analytical approximation of $|b_f|^2 + p_\uparrow$ (lines) for the parameter values $\epsilon = 0.1$, $\beta\Delta E_A = 10$, $g^2 = 10^{-3}$ (upper line and symbols) and $g^2=0$ (lower line and symbols).}
    \label{return}
\end{figure}

The approximations above are shown together with full master equation solutions in Fig. \ref{return}, again for $\epsilon=0.1$ and $\beta \Delta E_A =10$ for the ramp X $\rightarrow$ A as a function of the sweep rate $v_R$ for two cases of $g^2=10^{-3}$ and $g^2 =0$. The predictions based on the analytical estimates, i.e. $1-\rho_{gg,f} = |b_f|^2 + p_\uparrow$ are shown by black and blue lines for the two values of $g^2$, respectively. The corresponding solutions based on the master equation are shown by solid symbols for the two values of $g^2$ as well. We see that the approximations are relatively accurate in this regime. The decline for $v_R \rightarrow 0$ of the master equation data set for $g^2=10^{-3}$ (black dots) is due to relaxation back from the excited state, the effect ignored in the analytical estimates.

\begin{figure}
    \begin{center}
    \includegraphics[scale=.26]{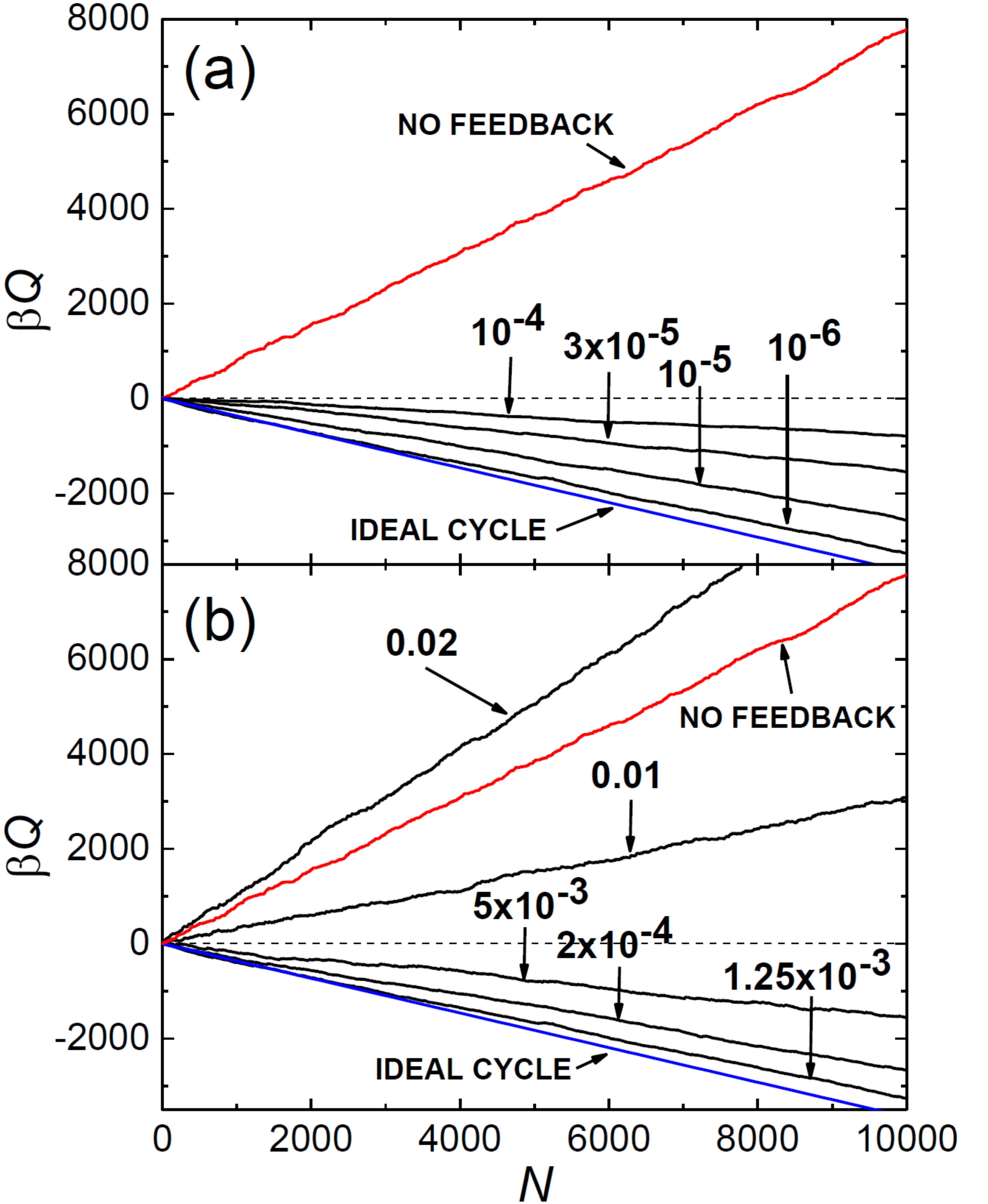}
    \end{center}
    \caption{Results of stochastic quantum trajectory simulations over $N$ cycles of demon operation. In (a) the quasistatic ramp rate $v$  (with $v_R = 1.25\cdot 10^{-3}$) and in (b) the return ramp rate $v_R$ (with $v = 1\cdot 10^{-6}$) are varied as indicated by the numbers with arrows. The top (red) line in both (a) and (b) shows the result when the demon is not active, whereas the bottom (blue) line shows the result of elusive ideal cycles. The horizontal dashed line shows the break-even result.}
    \label{cycles}
\end{figure}

As the last step of the numerical analysis of the demon, we have performed simulations based on the quantum trajectory method in cyclic operation. In this situation, once a cycle has finished, the state of the qubit when reaching point A serves as a new initial state of the next adiabatic ramp A $\rightarrow$ X. Thus, dynamics of the previous return path makes the initial state not necessarily the ground state, which leads to extra dissipation. Figure \ref{cycles} shows results of this analysis, where heat accumulation over $N$ cycles up to $N=10^4$ has been analyzed under different driving parameters. In all these simulations, $\epsilon = 0.1$, $\beta\Delta E_A =10$, and $g^2=10^{-3}$. In Fig. \ref{cycles} (a), the rate of return sweep remains constant, at $v_R = 1.25\cdot 10^{-3}$, which is close to its optimum value. The parameter indicated within the figure is the ramp rate $v$ of the adiabatic leg. We see clear cooling by the MD and correspondence with Fig. \ref{nonad}. For comparison, the fast rising (red) line indicates results of the calculation where the parameters are as for the lowest black curve in the figure, but without the measurement and $\pi$-pulse before the ramp back to A. This protocol is similar to a periodically driven two level system, which has been studied in detail recently \cite{Philip,Simone}. Naturally in such deterministic cycles there is positive heat dissipated into the bath according to the second law. The lowest (blue) line shows the result of Eq.~\eqref{qd4} which would be achieved if the first ramp were fully adiabatic, and the system would remain constantly in the ground state during the return leg. In Fig. \ref{cycles} (b) we show similar data as in (a), but now keeping the adiabatic ramp at a constant rate $v=10^{-6}$. The return rate $v_R$ is in turn varied here. We see correspondence with Fig. \ref{return}, showing the optimum rate of about $v_R=1.25\cdot 10^{-3}$.

It is possible to make some general statements as to the thermodynamic efficiency of the considered quantum MD. Thermodynamic cost 
of information acquired by the detector, based on Landauer's principle, is $-k_BT(\rho_{gg}\ln\rho_{gg} +\rho_{ee}\ln\rho_{ee})$, where $\rho_{ee}=1-\rho_{gg}$.  For ideal cycles considered above, this cost equals the heat $|\langle Q\rangle |$ of Eq. \eqref{qd4} extracted in a 
cycle. In this respect, ideally, the presented MD is fully reversible. Work $W_\pi$ done in applying the $\pi$-pulse can obtain two values, $W_\pi=0$ if the system is in the ground state and nothing is done, or $W_\pi=-\Delta E_0$ if the two-level system is in the excited state 
and the pulse is indeed applied. Therefore, averaging over many cycles, we obtain
\begin{equation} \label{ba11}
\langle W_\pi\rangle = -\rho_{ee}\Delta , E_0= - \frac{\Delta E_0}{1+e^{\beta \Delta E_0}}.
\end{equation}
Inserting Eq.~\eqref{qd4} for average heat, we find that
\begin{equation} \label{ba3}
\langle W_\pi \rangle -\langle Q\rangle = \beta^{-1}\ln (1+e^{-\beta \Delta E_0}) > 0.
\end{equation}
Both $\langle W_\pi \rangle$ and $\langle Q\rangle$ are negative. Thus the work extraction in this step is less than the heat extraction in the whole cycle. In steady state, the average work in the whole cycle $\langle W \rangle = \langle Q \rangle$ according to Eq.~\eqref{na1}, see also \cite{suppl}.

Finally, a few words about experimental feasibility of the proposed qubit demon. The principle studied here applies naturally to a large variety of systems, including solid-state qubits. As a concrete example, we may consider a superconducting qubit, e.g., a charge qubit \cite{averin96,nakamura99} or transmon qubit \cite{nakamura99,koch07}. Typical energy scale in this case is $\Delta E_A/k_B \sim 1$ K, as aluminium, the most common superconductor in this respect, has an energy gap $\Delta/k_B \approx 2$ K. Since the operating (bath) temperature of these devices is $T\sim 100$ mK, we indeed have $\beta \Delta E_A \sim 10$, as in the numerical examples above. In a charge-type qubit, $\epsilon$ is the ratio of the Josephson coupling and capacitive charging energies, and can be made to be $\sim 0.1$, based on design and tuning by magnetic flux. The value of $g^2 = 10^{-3}$ would correspond to a relaxation time of the qubit, $T_1=1/\Gamma_\Sigma$, which is then in the range
$0.1 ... 1$ $\mu$s,
quite feasible for the current technology. The ramp times for the demon fall also in a favourable regime. Since $\tau \sim 10^5 \hbar /\Delta E_A \sim 1 $ $\mu$s yields still about 80\% efficiency in the adiabatic leg as compared to the infinitely slow ramp (see Fig. \ref{nonad}), we may take this as a naive trade-off value in terms of efficiency vs. power extraction. The duration of the whole cycle is determined almost entirely by $\tau$. It is interesting to see that 1 $\mu$s is about 100 times faster than a typical electron-phonon relaxation time at these low temperatures \cite{taskinen04,gasparinetti15}. Therefore, if the bath is composed of, e.g., a micron-scale metallic calorimeter, it is reasonable to expect that the steady-state temperature of this absorber would decrease once the MD is set into cyclic operation \cite{gasparinetti15}. Measuring this temperature drop would serve as a smoking gun of the proof of the principle of the demon. Last, we found that $\tau_R \sim 10^3 \hbar/\Delta E_A \sim 10 $ ns would be optimal with these parameters of the MD. This again is fully compatible with the time scales of the gate and flux operations of superconducting qubits.

The work has been supported by the European Union FP7 project INFERNOS (grant agreement 308850) and by the Academy of Finland (projects 272218 and 284594). D.V.A. was also supported in part by the NSF grant PHY-1314758.

\section{Appendix I: Suppression of the Landau-Zener processes by smooth drive}

In this Section of the Supplementary Material, we provide a brief discussion of the effect of smooth drive on the amplitude
$b_f$ of the Landau-Zener excitation of the qubit out of the ground state in the process of rapid ramp back from the level-crossing
point to the initial state. The most basic ramp-back process studied in the main text is characterized by the time dependence of the
control parameter $q(t)$, which is continuous itself, but has discontinuous first derivative: $q(t) \equiv 0$ for $t<0$ (we take the
instant of measurement and, when needed, the $\pi$-pulse, to be $t=0$) and $q(t) = \dot{q}t=t/(2\tau_R)$ for $t>0$. This discontinuity
of the first derivative leads to the excitation amplitude $b_f$ which decreases as $1/\tau_R$ with the ramp time $\tau_R$ -- see
Eq.~(10) of the main text. The point of our discussion here is to demonstrate that the amplitude $b_f$ can be made to decrease
faster with the ramp time $\tau_R$, if any discontinuities of the time dependence $q(t)$ of the control parameter at the point $t=0$
where the ramp is switched on, are suppressed more strongly. The statement that we want to make here, demonstrated by examples, is that
the amplitude $b_f$ can be made to decreases as $1/\tau_R^n$, if the switching function $q(t)$ is such that only its $n$th-order
derivative is discontinuous, while all previous derivatives are continuous, and exponentially, if $q(t)$ has continuous derivatives
of any order.

To do this, we need to calculate the excitation probability $|b_f|^2$ for evolution starting at the level-crossing point.
For the part of the MD cycle, ramp back to the initial state, which is considered in this discussion, we are interested in the 
regime of sufficiently slow time evolution, when the probability $|b_f|^2$ is small, i.e., the qubit stays predominantly in the 
ground state. On the other hand, this ramp of the control parameter $q(t)$ should be fast enough for the thermal-bath-induced 
transitions to be negligible in the calculation of $b_f$. As follows from Eq.~(6) of the main text, in this regime, the evolution 
equation of the excitation amplitude $b$ is
\begin{equation}
\dot b = \frac{1}{2}\Omega e^{+i\int_0^t \Delta E(t')dt'/\hbar} ,
\label{SM1} \end{equation}
and can be integrated to give the following expression for the amplitude at the end of the ramp:
\begin{equation}
b_f = \int_0^{1/2}\frac{(\epsilon/2)dq}{q^2+\epsilon^2}\exp \{\frac{2i\Delta E_A}{\hbar}\int_0^q
\frac{dp}{\dot{p}}(p^2+\epsilon^2)^{1/2}\} \, .
\label{SM2} \end{equation} 
As discussed above, we are interested in the time evolution which is adiabatic with respect to the Landau-Zener transitions. 
Quantitatively, this means that the parameters in Eq.~\eqref{SM2} should satisfy the conditions $1\gg \epsilon \gg \hbar/(\Delta E_0 
\tau_R)$. In this case, the first of these inequalities ensures that the integral \eqref{SM2} converges at $q \ll 1$, and can be 
rewritten in terms of $x=q/\epsilon$ as follows: 
\begin{equation}
b_f = \frac{1}{2}\int_0^{\infty}\frac{dx}{x^2+1}\exp \{\frac{i\Delta E_0}{\hbar}\int_0^x \frac{dy}{\dot{y}} 
(y^2+1)^{1/2}\} \, .
\label{SM3} \end{equation}
Note, however, that by extending the limit of integration to infinity, we are neglecting possible contribution to $b_f$ from 
the discontinuity in the control parameter drive at the switching-off point $t=\tau_R$. In principle, the time dependence of 
$q$ at that point also needs to be smoothed out in the same way as at the switching-on point $t=0$ we are discussing. 

Equation \eqref{SM3} is valid for any time dependence of the control parameter $q(t)$ leading from $q=0$ at $t=0$ to $q=1/2$ at 
$t=\tau_R$. To obtain specific results, we first consider $q(t)$ which satisfies this requirement, and at the same time, has a 
discontinuous $n$th derivative at $t=0$:
\begin{equation}
q(t)= t^n/(\tau_R^n+t^n) \, ,\;\;\;\; t>0\, .
\label{SM4} \end{equation}
Since only the small-$q$ part of the time dependence $q(t)$ affects the integral \eqref{SM3}, i.e., the part at $t\ll \tau_R$, 
Eq.~\eqref{SM4} can be inserted into Eq.~\eqref{SM3} in the simplified form, $q(t)=(t/\tau_R)^n$. Then $\dot{y}(t)=nt^{n-1}/
(\epsilon \tau_R^n)$ and $\dot{y}(y)=ny^{1-1/n}/(\epsilon^{1/n} \tau_R)$. With this expression, the integral \eqref{SM3}
takes the form: 
\begin{eqnarray}
b_f = \frac{1}{2}\int_0^{\infty}\frac{dx}{x^2+1}\exp \{\frac{i\Delta E_0\tau_R \epsilon^{1/n}}{n\hbar}\nonumber \\
\times \int_0^x dyy^{-(1-1/n)} (y^2+1)^{1/2}\} \, ,  \label{SM5} \end{eqnarray}
and can be transformed further by the substitution $u=(\epsilon y)^{1/n}$, $v=(\epsilon x)^{1/n}$:
\begin{eqnarray}
b_f = \frac{n}{2\epsilon} \int_0^{\infty}\frac{dv v^{n-1}}{(v^{2n}/\epsilon^2)+1} \exp \{\frac{i\Delta E_0\tau_R}{\hbar}\nonumber \\
\times \int_0^v du[(u^{2n}/\epsilon^2)+1]^{1/2}\} \, .  \label{SM6} \end{eqnarray}
The integration region in this equation (real half-axis of $v$) can be rotated into the imaginary half-axis, demonstrating  
explicitly that in the adiabatic limit, when $\epsilon^{1/n}\gg \hbar/\Delta E_0\tau_R$ (for $n=1$, this is the same condition that 
ensures that the usual Landau-Zener probability is small), the integral is converging at $v\ll \epsilon^{1/n}$, and can be simplified 
and evaluated accordingly:
\begin{equation}
|b_f| = \frac{n}{2}\int_0^{\infty} dv v^{n-1} e^{-\Delta E_0 \tau_R  v/\hbar} = \frac{n!}{2} \Big(\frac{\hbar}{\Delta E_0 \tau_R}\Big)^n. 
\label{SM7} \end{equation}
We see that, indeed, the discontinuity of the $n$th order derivative of the time dependence of the control parameter $q(t)$ at the 
switching-on point $t=0$ implies that the amplitude $b_f$ of the qubit transition into the excited state in the ramp-back part of the 
MD cycle is suppressed with the increasing ramp time $\tau_R$ as $1/\tau_R^n$. 

This implies that one can make the the amplitude $b_f$ to decrease exponentially with $\tau_R$, if the switching-on process $q(t)$ has 
continuous derivatives of any order. Since $q(t)\equiv 0$ for $t<0$, this can happen only if $q(t)$ is non-analytic at $t=0$. 
As an example, we take 
\begin{equation}
q(t)= \frac{1}{2} e^{1-\tau_R/t} \, ,\;\;\;\; t>0\, .
\label{SM4} \end{equation} 
Estimating the integral of the evolution equation \eqref{SM1} for this $q(t)$ with the exponential accuracy by the steepest-decent 
method, we find
\begin{equation}
|b_f| \simeq e^{-(2\Delta E_0 \tau_R/\hbar)^{1/2}} .
\label{SM8} \end{equation}
This example shows explicitly that the excitation amplitude $b_f$ can be made exponentially small for smooth switching-on process 
of the control parameter ramp.
\begin{figure}
    \begin{center}
    \includegraphics[scale=.24]{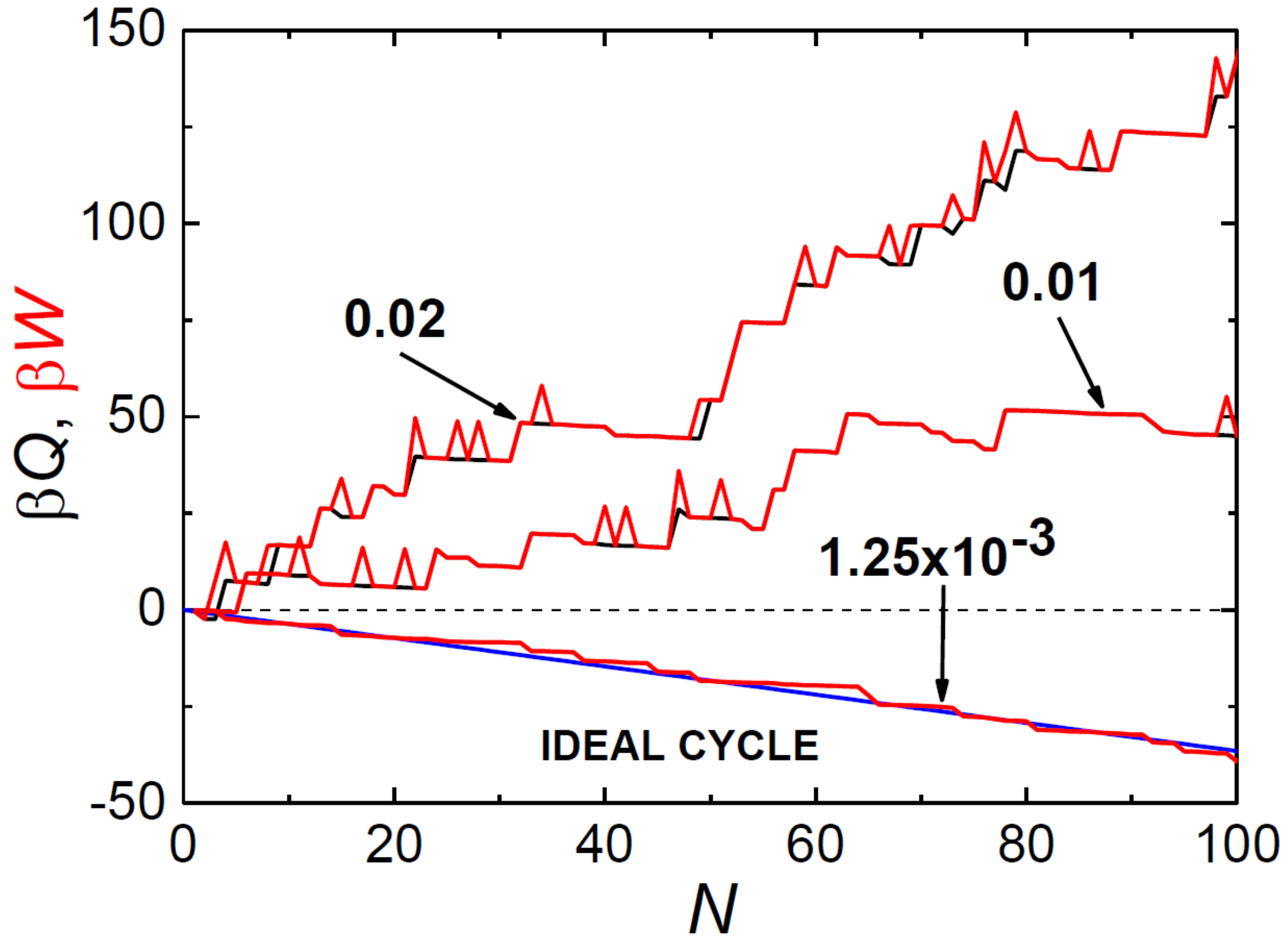}
    \end{center}
    \caption{Results on $Q$ (black line) and $W$ (red line) of stochastic quantum trajectory simulations over $N$ cycles of demon operation. The quasistatic ramp has the rate $v = 1\cdot 10^{-6}$, and the return ramp rate $v_R$ is varied as indicated by the numbers with arrows. The bottom (blue) line shows the result of elusive ideal cycles. The horizontal dashed line shows the break-even result. The other parameters in the simulation are $\epsilon = 0.1$, $\beta\Delta E_A = 10$, and $g^2 = 10^{-3}$. Since temperature is rather low, the system happens to start in the ground state in all the three presented examples, and therefore $W$ either coincides with $Q$ or exceeds it by $\Delta E_A$ everywhere.}
    \label{QandW}
\end{figure} 
\\
\section{Appendix II: Work and heat in cyclic operation of the device}
To substantiate our general claim that in cyclic operation work and heat per cycle are equal on average, $\langle W\rangle = \langle Q\rangle$, we have analyzed in Fig \ref{QandW} numerically the cumulative work and heat by the same method as presented in Fig. 4 of the main text, but now only over $N=100$ cycles. The difference between the two quantities is the energy stored in the qubit as compared to its initial value when the operation starts. In the long time limit ($N\rightarrow \infty$), the averages of the two quantities per cycle coincide.

One should bear in mind that the data in Fig. \ref{QandW} cannot be interpreted as a result of many measurements of work in a single trace shown. It rather presents the evolution of $Q$ extracted from the bath. Then, assuming this heat trajectory, work is measured after the $N$:th cycle. So the work values represent a likely outcome of the measurement along the trajectory, if $W$ was measured at this point to terminate the protocol at the given value of $N$.

\end{document}